\documentclass[aps,prd,twocolumn,showpacs,superscriptaddress,nofootinbib]{revtex4-1}

\usepackage[pdftex]{graphicx}

\usepackage{amsmath}
\usepackage[latin1]{inputenc}
\usepackage{txfonts}
\usepackage{hyperref}

\usepackage{xcolor}

\begin{document}

\title{Symmetries and synchronization in multilayer random networks}
\author{Alberto Saa} 
\email{asaa@ime.unicamp.br}
\affiliation{
Department of Applied Mathematics, 
 University of Campinas,  13083-859 Campinas, SP, Brazil.}

\date{\today}

\begin{abstract}
In the light of the recently proposed scenario of 
asymmetry-induced synchronization (AISync),  in which dynamical
uniformity and consensus in a distributed system would demand certain asymmetries in the underlying network,  
we investigate here the influence of some   regularities in the interlayer connection patterns  on the
synchronization properties of multilayer  random networks. More specifically,  by considering a Stuart-Landau model of complex oscillators with random frequencies,
we   report for  multilayer networks a dynamical behavior that could be also classified as
a manifestation of AISync. We show, namely, that the presence of certain symmetries in the interlayer connection pattern  
 tends  to diminish the synchronization capability of the whole network or, 
in other words, asymmetries in the interlayer connections would enhance synchronization in such structured networks. Our
results might help the understanding not only 
of the AISync mechanism itself, but also its possible role in the determination 
of the interlayer connection pattern of multilayer and other structured networks with optimal
synchronization properties.
 \end{abstract}

\maketitle

\section{Introduction}

Recently, Nishikawa and   Motter \cite{NishMotter} put forward a rather intriguing dynamical phenomenon present in networks of coupled oscillators: the asymmetry-induced symmetry, or
asymmetry-induced synchronization (AISync), a scenario where the
occurrence of  dynamical uniformity and consensus  does require certain asymmetries in the underlying network of connections among the oscillators. In many senses,
this new phenomenon could be considered as the converse of the well known symmetry breaking process, in which  some solutions or the dynamical behavior   of a given system do  not inherit all the symmetries of the respective governing equations\cite{NishMotter}.
A general scheme for building networks exhibiting this novel dynamical behavior was just introduced in \cite{ZhangNishMotter}, where the authors also argue that the asymmetry-induced synchronization  
should be a rather prevailing   behavior in multilayer networks of coupled oscillators. 
For a recent exposition on AISync, see \cite{ZhangMotter}
 Besides its unequivocal interest from a more fundamental point of view, the AISync
  has also a wide range of possible applications, particularly in synchronization problems \cite{ pikovsky2001,arenas2008}, where the new phenomenon was indeed initially discovered.   
The AISync challenges frontally the idea that synchronization states should be  promoted by symmetric configurations. Nevertheless, there are indeed plenty of relevant examples of synchronization
driven by symmetries in networks of oscillator, see, for instance, 
\cite{nicosia2013,pecora2014,sorrentino2016,cho2017}.

Here, we report some new results which may  also be classified as
a manifestation of AISync. 
Differently from \cite{NishMotter}, where the analysis was performed for small scale regular networks, we consider spontaneous synchronization in large scale  random networks. More specifically,  in the same line of \cite{ZhangNishMotter}, we consider  complex oscillators  on 
multilayer random networks\cite{multi1,multi2} and, 
 by exploring some recent  ideas and algorithms on optimal synchronization\cite{optimal}, we show that 
the presence of
certain regularities
in the interlayer connection pattern      tends to diminish the synchronization capability of the coupled oscillators system or, in other words,    asymmetries 
in the connection between layers would enhance synchronization in these kind of structured  networks. In this paper, we   call   structured random network a  random multilayer network where inlayer and interlayer connections can have different statistical properties, which could mimic  real situations where the layers and the connection among them  evolve and are selected differently.

Following the same principles of   \cite{NishMotter} and \cite{ZhangNishMotter} , we consider both phase and amplitude
effects by studying the so-called Stuart-Landau (SL) model with complex oscillators on  an $N$-nodes   random network   
\begin{equation}
\label{SL}
\dot z_k = \left( \alpha^2 + i\omega_k - |z_k|^2\right) z_k + \lambda \sum_{j=1}^Na_{kj}(z_j-z_k),
\end{equation}
where $z_k$ is the (complex) state of   the oscillator located at the $k$ node, with $\omega_k$ standing for its natural frequency, which we also assume to be  a random variable.  
The entries $a_{kj}$ correspond here to the usual adjacency matrix   for undirected and unweighted networks, and $\lambda$ defines the (real and uniform) coupling strength among the complex oscillators. The real parameter $\alpha$ determines the stability properties of the limit cycle $|z_k|^2 = \alpha^2$, which is clearly present for $\lambda=0$.
 For larger values of $\alpha^2$ (compared with $\lambda$), one recovers the paradigmatic Kuramoto model \cite{kuramoto1975,strogatz2000,acebron2005}.
 For further properties and references on the SL model, see \cite{SL1,SL2,SL3}, for instance.
In our simulations, we will use the real version of the Eq. (\ref{SL}). Introducing $z_k=\rho_ke^{i\theta_k}$, we have  
\begin{equation}
\dot\rho_k + i\rho_k \dot\theta_k = \left( \alpha^2 - i\omega_k - \rho_k^2 \right)
\rho_k + \lambda \sum_{j=1}^N a_{kj}\left(\rho_je^{i(\theta_j-\theta_k)} - \rho_k \right).
\end{equation}
Collecting the real and imaginary parts, one has
\begin{eqnarray}
\label{SLR1}
\dot\rho_k &=& \left( \alpha^2 -  \rho_k^2 \right)\rho_k -\lambda 
\sum_{j=1}^N {\ell}_{kj} \rho_j \cos \left(\theta_j-\theta_k \right),\\
\label{SLR2}
\rho_k\dot\theta_k &=& -\omega_k\rho_k + \lambda  \sum_{j=1}^N 
a_{kj}\rho_j \sin \left(\theta_j-\theta_k \right),
\end{eqnarray}
where $\ell_{kj}$ stands for the usual network Laplacian matrix components.

It is worth to mention that the self-interacting term in (\ref{SL}) corresponds  to the generic behavior near a Hopf bifurcation involving limit cycles, and thus such model is also widely known as Andronov-Hopf oscillators in the literature, see \cite{Izhi}, for instance.
Since we are mainly interested in synchronization properties, a
 convenient description for the global state of the SL model will be given by the order parameter $r$ defined as
\begin{equation}
r(t)  =  
\left|\frac{1}{N}\sum_{j=1}^{N}e^{i\theta_j}\right| 
.
\label{order_parameter}
\end{equation}
Clearly, the behavior of the order parameter $r$, which depends only on the oscillator phases,    is analogous to the Kuramoto
case: $r \approx N^{-1/2}$ for incoherent motion, whereas $r \approx 1$ for a fully synchronized 
state. 

\begin{figure}[t]
\includegraphics[scale=0.4]{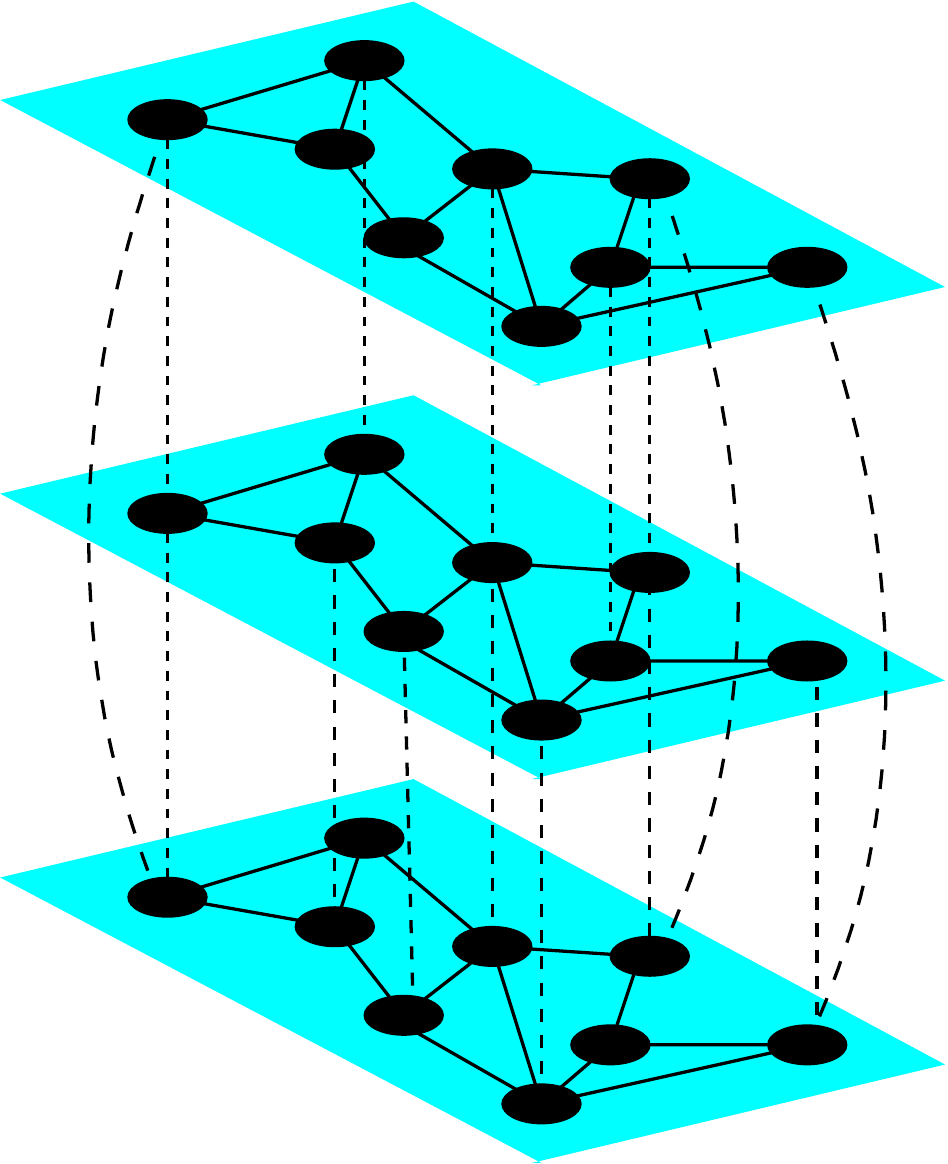} 
\includegraphics[scale=0.4]{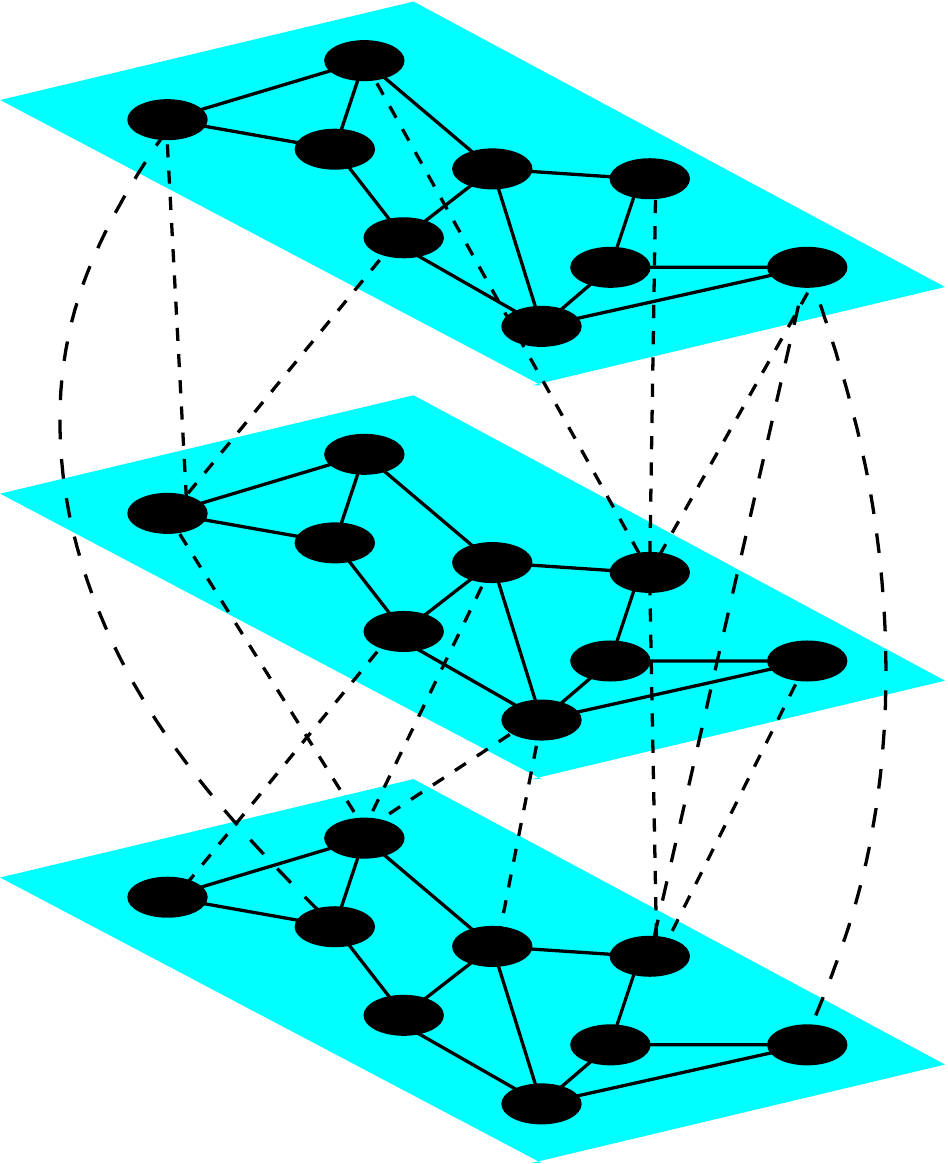} 
\caption{Typical structured networks considered in this paper. Left: a trilayer random network  with   identical layers and a {\em diagonal} interlayer connection pattern, {\em i.e},   connections between the layers are fulfilled  only by {\em equivalent} nodes in each layers. The diagonal pattern corresponds, probably, to the most regular network topology one might consider in this context.   Right: a trilayer random network  with   the same
layers  of the previous case, with also the same number of interlayer connections, but with a non-diagonal pattern. 
The network with diagonal interlayer connections may exhibit  some  discrete symmetries which   are always absent for the non-diagonal case,
see Section III for the 
 precise definition and meaning of such
symmetries. 
We have shown that the regular topology (Left)
has always impaired  synchronization capabilities  if compared with the asymmetric case (Right). 
}
\label{Fig1}
\end{figure}

We will consider in this paper the case of multilayer random networks with identical
layers, see Fig. \ref{Fig1} for some typical examples. 
The interlayer connection pattern 
in these networks can exhibit or not some discrete
symmetries corresponding to entire layer permutations. The precise definition and meaning of such
symmetries will be discussed in Section III.  The most regular connection patterns   one may consider in this context are probably those ones with  {\em diagonal} interlayer connections, see Fig. \ref{Fig1}, in the sense that the connection among layers is fulfilled  only by {\em equivalent} nodes in each layer.  
 We will show  that    multilayer  networks with 
such  regular interlayer connection pattern 
   tend to have impaired synchronization capabilities  if compared with similar networks without these regularities 
 in the interlayer connections.
  Typically, starting with   a regular multilayer network, its synchronization properties can be considerably enhanced by applying our optimization rewiring algorithm introduced
  in \cite{optimal} for the
  interlayer connections if 
one abandons  the diagonal  pattern. This is, indeed, our main result: 
the breaking of the interlayer connection symmetries will lead to the enhancement of the network synchronization capabilities,  in the same spirit of  Motter-Nishikawa   AISync \cite{NishMotter}. We   also corroborate one of the main   conclusions of \cite{ZhangNishMotter}, namely that  AISync  should be a rather generic property of structured networks, {\em i.e.}, 
unstructured networks do not exhibit in general such an anti-correlation between symmetries and synchronization. 

In the next section, we will revisit the rewiring  algorithm introduced in \cite{optimal}, based
on the Gottwald dimensional   reduction approach  proposed in \cite{gottwald2015}, for
the optimization of synchronization in the  Kuramoto model. Basically, we will show that 
the same ideas behind the 
 Gottwald dimensional reduction and the optimization algorithm   can be conveniently 
  employed  also for the
 SL model. We also derive some new mean-field approximation  formulas which have proved to be  useful  
 in estimating  the efficiency of our algorithm.
 In Section III,  we will consider the synchronization capabilities of 
 large scale multilayer random networks with different interlayer connection patterns, and
 our main result will be established, with the  predictions of the mean-field analysis being confirmed  
 by exhaustive numerical simulations.  
 The  last section is devoted to some concluding remarks. We discuss, in particular, the differences between structured and non-structured networks and why the very idea of asymmetry-induced synchronization  is much more relevant to the former.

\section{Dimension  reduction  for complex oscillators}

In \cite{optimal},  we introduced a  rewiring algorithm for 
the optimization of synchronization in
the Kuramoto model by exploring  the dimensional
reduction approach recently proposed by Gottwald
\cite{gottwald2015}, which, on the other hand, is based in the introduction of a 
collective coordinate for the Kuramoto oscillators  in the same spirit of the Ott-Antonsen
ansatz \cite{ott2008,ott2009}. The extension  of this optimization scheme  
for the SL model
will be
instrumental in the present analysis. 
  The  Gottwald collective ansatz \cite{gottwald2015} can be adapted for the  present case  as
\begin{equation}
\label{ansatz}
z_k(t) = \rho_k(t) e^{i\left(\omega_k\beta(t) + \Omega t + \varphi_0\right)},
\end{equation}
which will be able to capture all the essential properties of the synchronized states for the
SL model. The offset parameter $\varphi_0$ can be conveniently set to zero without loss of generality, since the SL model   is an autonomous system and $\varphi_0$ can be easily absorbed by a simple shift in $t$.   From the equations of motion (\ref{SLR1}) and
(\ref{SLR2}), one has the following dynamical equations for the real variables  $\rho_k$  and $\beta$
\begin{equation}
\label{eq1}
\dot\rho_k      = F_k(\rho_j,\beta) = 
\left(\alpha^2    -\rho_k^2\right)\rho_k    -
\lambda \sum_{j=1}^N\ell_{kj} \rho_j \cos\beta(\omega_j-\omega_k) , 
\end{equation}
  and
\begin{equation}
\label{Im}
  \rho_k \omega_k\dot\beta      = 
  (\omega_k-\Omega)  \rho_k    +
\lambda \sum_{j=1}^Na_{kj} \rho_j \sin\beta(\omega_j-\omega_k) .
\end{equation}
Ir order to cast 
Eq. (\ref{Im})   in a more convenient form, we multiply both sides by $\rho_k\omega_k$ 
\begin{equation}
  \rho_k^2 \omega_k^2\dot\beta      = 
   \rho_k^2\omega_k^2 -\Omega   \rho_k^2\omega_k    +
\lambda \sum_{j=1}^Na_{kj} \omega_k\rho_k\rho_j \sin\beta(\omega_j-\omega_k)  .
\end{equation}
Now, summing over $1\le k\le N$, dividing by $N$ and rearranging  the terms, one gets
\begin{eqnarray}
\label{beta}
\dot\beta  = G(\rho_j,\beta) &=& 1 -\Omega\frac{\left\langle\rho_k^2\omega_k\right\rangle}{\left\langle\rho_k^2\omega_k^2\right\rangle}   \\
& &  
+\frac{\lambda}{N\left\langle \rho_k^2\omega_k^2\right\rangle } \sum_{k =1}^N \sum_{j =1}^N a_{kj} \omega_k\rho_k\rho_j \sin\beta(\omega_j-\omega_k), \nonumber
\end{eqnarray}
where the brackets denote simple averages as, for instance, in
\begin{equation}
\left\langle\rho_k^2\omega_k^2\right\rangle = \frac{1}{N}\sum_{k=1}^N\rho_k^2\omega_k^2.
\end{equation}
In contrast with the case of the Kuramoto model considered in \cite{optimal}, one cannot assure, in principle, that the rigid rotation $\Omega$ vanishes for a synchronized state in networks with a  frequency distribution  $g(\omega)$ with null average, for instance.  In fact, for  a fixed point  $\left(\bar\rho_j,\bar\beta\right)$  of (\ref{eq1}) and (\ref{beta}), we will have
\begin{equation}
\label{Omega}
\Omega = \frac{\left\langle \bar\rho^2_k\omega_k\right\rangle}{\left\langle \bar\rho_k^2\right\rangle}.
\end{equation}
Notice that  for uniform $\bar\rho_k$ (the Kuramoto limit), $\Omega$ does indeed vanish if 
$\langle \omega_k \rangle =0 $ and, moreover, in this case equation (\ref{beta}) will coincide with the Gottwald 
dimensionally reduced 
 equation introduced in  \cite{gottwald2015} for the Kuramoto model. Hereafter, unless explicitly stated otherwise, we will assume  we are   
 dealing with symmetric distributions $g(\omega)$ with null average. 

From our ansatz (\ref{ansatz}), it is clear that 
the fixed points $\left(\bar\rho_j,\bar\beta\right)$ of (\ref{eq1}) and (\ref{beta})      correspond     to synchronized states.
    Furthermore, for small values of $\bar \beta$, they will typically be 
     optimally synchronized states, in the sense that they  will exhibit  an order parameter $r$ given by (\ref{order_parameter}) close to 1. 
For large scale networks, one can estimate  $r$ for a given fixed point,   with good accuracy,  by using a mean-field approach.  
Directly from (\ref{order_parameter}) and (\ref{ansatz}), we have   
\begin{equation}
\label{rfixed}
r(\beta) = \left\langle \cos \beta\omega_k \right\rangle = \int d\omega \, g(\omega) \cos  \beta \omega  
\end{equation}
for a symmetric    distributions $g(\omega)$ of frequencies with 
null average. The most commonly used distributions in the literature are the normal and the homogeneous, which will correspond, respectively, to the following mean-field expressions for $r$
\begin{equation}
\label{rn}
r_{\rm n}  =
\exp\left( -\frac{1}{2} \beta^2   \sigma_\omega^2\right)
\end{equation}
and
\begin{equation}
\label{ru}
r_{\rm u} = \frac{\sin \sqrt{3 } \beta \sigma_\omega}{\sqrt{3}  \beta\sigma_\omega}.
\end{equation}
For both case, we have
\begin{equation}
r\approx 1 - \frac{1}{2} \beta^2  \sigma_\omega^2
\end{equation} 
for small $\beta$.  Notice that  synchronized states with $r\approx 1$   require also 
a small frequency standard deviation 
$\sigma_\omega  = \sqrt{\left\langle \omega^2_k \right\rangle}$.

\begin{figure}[t]
\includegraphics[scale=0.6]{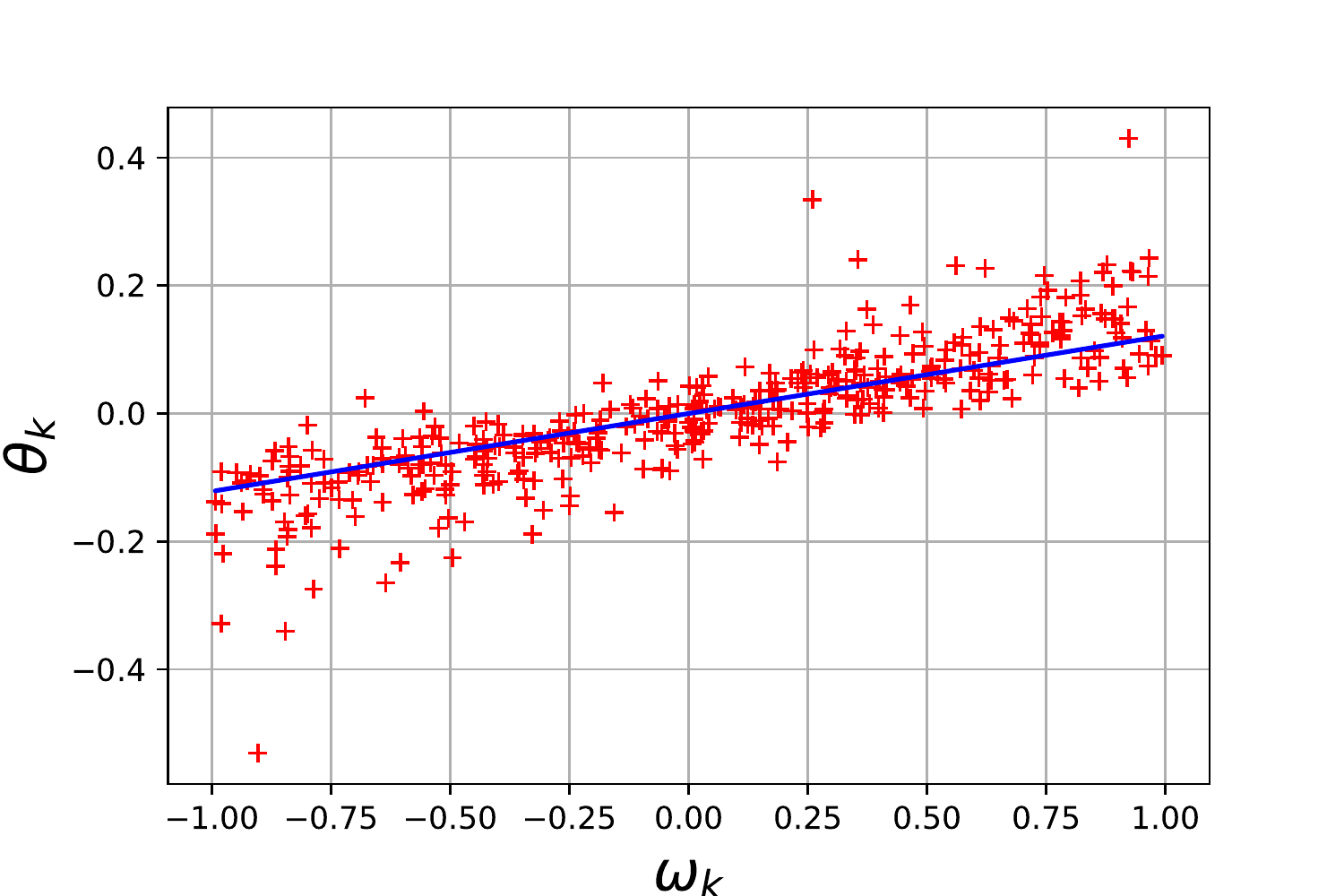} 
\includegraphics[scale=0.6]{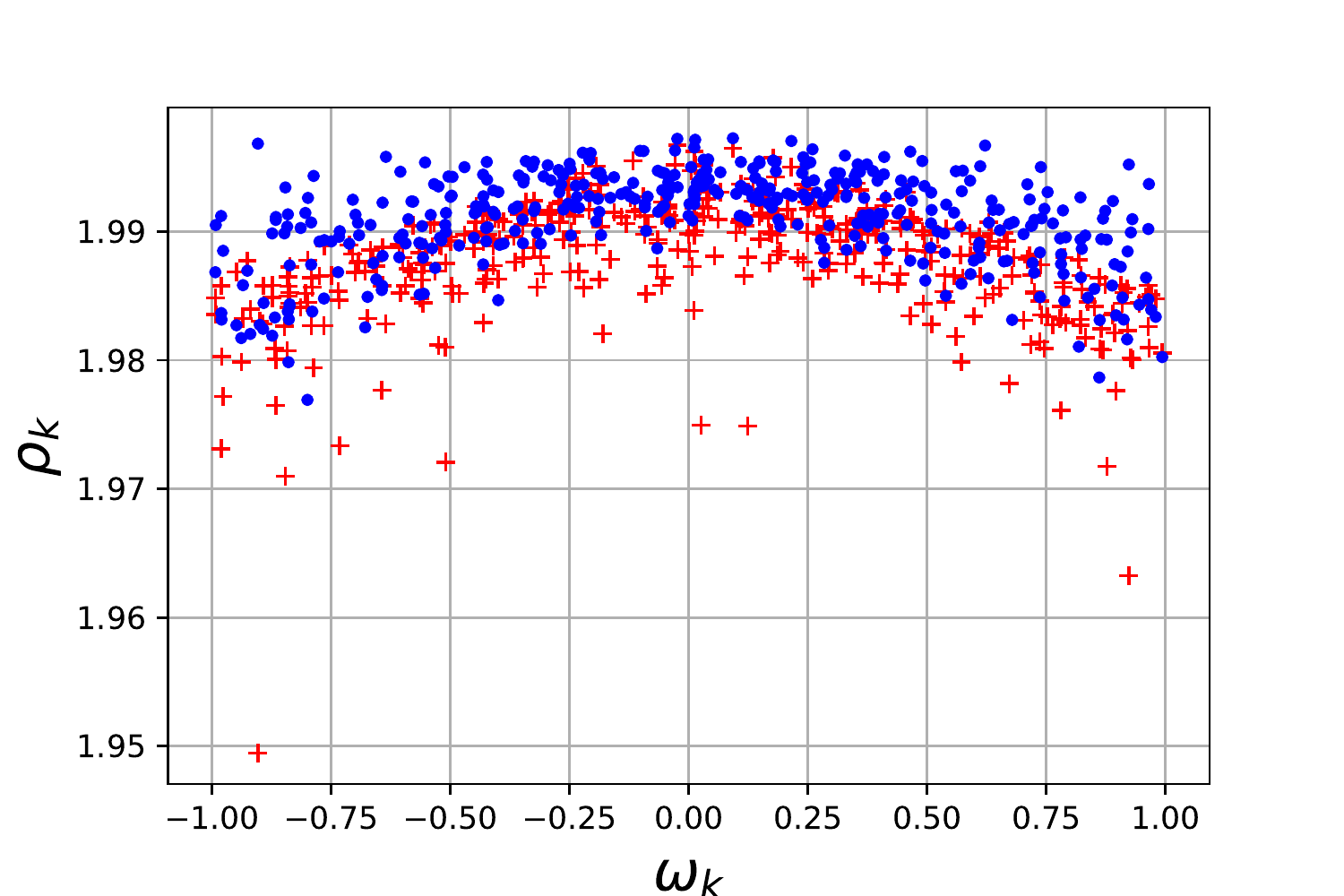} 
\caption{Typical asymptotic behavior of a synchronized state in the SL system (\ref{SL}), where $z_k = \rho_ke^{i\theta_k}$. The depicted cases correspond  to an Erd\"os-R\'enyi random network with 400 nodes,  with
average degree $\langle d_k \rangle= 8.26$, $\alpha = 2$, $\lambda = 1$, and $\omega_k$ draw from the interval $(-1,1)$ with uniform distribution. Top: the 400 points $(\omega_k,\theta_k(t))$  
 for  $t$ large enough to reach a stationary regime (namely in this case, $t=5$), starting from random initial conditions. The straight line has inclination given by the prediction (\ref{fixed-beta}), $\bar\beta = 0.12$, showing that the ansatz (\ref{ansatz}), which implies  
$\theta_k=\bar\beta \omega_k + \Omega t$ in the stationary regime, is indeed a good approximation. The rigid rotation in this case is $\Omega= 0.02$. Bottom: the red crosses are the points $(\omega_k,\rho_k(t))$ in the stationary regime, while the blue circles correspond to the prediction (\ref{fixed-rho}), confirming a good overall accuracy for our ansatz (\ref{ansatz}).  
 }
\label{Fig2}
\end{figure}

We are quite sure about the existence of the fixed points
 $\left(\bar\rho_j,\bar\beta\right)$ of (\ref{eq1}) and (\ref{beta})  
   for coupling constants $\lambda$ 
larger than some threshold value $\lambda_c$. In fact, we will   show   that we can even estimate $\lambda_c$ with good accuracy from the Gottwald dimensionally reduced approach.
The fixed points  $\left(\bar\rho_j,\bar\beta\right)$ correspond, of course, to the zeros 
 $F_k(\bar\rho_j,\bar\beta) =0$ and  $G(\bar\rho_j,\bar\beta) =0$. Since we are dealing with a system of non-linear   equations, we can have effectively  several fixed points, some of them might be even  dynamically stable. However, for our purposes, we will focus on the fixed point near $\rho_k=\alpha$ and $\beta=0$, which turns out to be always dynamically stable.  
Let us  consider a linear approximation for the system (\ref{eq1}) and (\ref{beta})
around  this point. By introducing $\rho_k = \alpha + \delta_k$, 
we have
\begin{eqnarray}
\dot\delta_k &=&  - 2\alpha^2\delta_k -\lambda\sum_{j=1}^N\ell_{kj}\delta_j, \\
\dot\beta &=& 1 - \lambda {\cal L} \beta -
\frac{2\Omega \left\langle \omega_k\delta_k\right\rangle}{\alpha  \left\langle \omega^2_k\right\rangle } ,
\end{eqnarray}
with
\begin{equation}
\label{calL}
{\cal L} = 
\frac{ \boldsymbol{\omega}^TL \boldsymbol{\omega}}{\boldsymbol{\omega}^T \boldsymbol{\omega}},
\end{equation}
where $L=[\ell_{kj}]$ is the usual Laplacian matrix for the network  and $\boldsymbol{\omega}$
is the $N$-dimensional vector formed by the oscillator natural frequencies, 
  $\boldsymbol{\omega}=[\omega_k]$. 
Since the Laplacian matrix is a non-negative diagonalizable matrix, $\lambda L +2\alpha^2 \boldsymbol{I}$   is invertible 
for $\alpha \ne 0$ and $\lambda >0$,  
 and hence the linearized fixed points are   $\bar\rho_k\approx\alpha$ and
 \begin{equation}
 \label{fixed-beta}
\bar\beta \approx \frac{1}{\lambda\cal L} ,
\end{equation}
which, as we will see below, are indeed  good approximations for $\left(\bar\rho_j,\bar\beta\right)$ for large values of $\lambda \cal L$. 
It is convenient for our purposes here to go one step further by expanding the equations (\ref{eq1})   up to second order, leading to the following second order approximation  in $\beta$ for the
fixed point $\bar\rho_k$
\begin{equation}
\label{fixed-rho}
 \bar\rho_k  \approx \alpha - \frac{ \lambda\bar\beta^2}{2\alpha } 
 \sum_{i=1}^N
  \sum_{j =1}^N  m_{ki} a_{ij}   (\omega_j-\omega_i)^2 ,  
\end{equation} 
where 
\begin{equation}
[m_{ki}] = \left(\frac{\lambda}{\alpha^2}   L+2  \boldsymbol{I}\right)^{-1}.  
\end{equation}
Also, from (\ref{fixed-rho}), we have the following approximation for the 
synchronized state 
rigid rotation
$\Omega$ given by (\ref{Omega})
\begin{equation}
\label{OmegaApprox}
\Omega \approx \langle \omega_k\rangle - \frac{\lambda\bar\beta^2}{N\alpha^2}  \sum_{k=1}^N\sum_{i =1}^N \sum_{j =1}^N m_{kj}a_{ij} \omega_i (\omega_j-\omega_i)^2.
\end{equation}
Notice  that for large $\alpha$ we get from (\ref{fixed-rho}), as expected, the Kuramoto limit of uniform amplitudes, $\bar\rho_k = \alpha$, for which $\Omega$ vanishes for $\langle \omega_k\rangle=0$. 
We have performed  exhaustive numerical simulations and could verify  the approximations (\ref{fixed-beta}) and
 (\ref{fixed-rho}) with good accuracy, see Fig. \ref{Fig2} for some typical results. 
Also, 
it is worth noting that  $\rho_k=0$, which of course corresponds to $z_k=0$ in the
original SL equations (\ref{SL}), irrespective of the value of $\beta$, correspond to fixed points in our system. 
Such (continuous) family of fixed points are rather spurious for our purposes  since they not correspond effectively 
to synchronized states. However, since
\begin{equation}
\frac{\partial F_k}{\partial \rho_j} = \alpha^2 \delta_{kj} - \lambda \ell_{kj}\cos\beta(\omega_j-\omega_k)
\end{equation}
at $\rho_k=0$, we see that such points can in principle be attractive for sufficiently 
large values of $\lambda$ and certain ranges of $\beta$. This is a point to keep in mind,
excessively  large values of $\lambda$ can
push the system into this attraction basin, jeopardizing the possibility of attaining 
eventually 
a synchronized state using the ansatz (\ref{ansatz}).

\begin{figure}[t]
\includegraphics[scale=0.6]{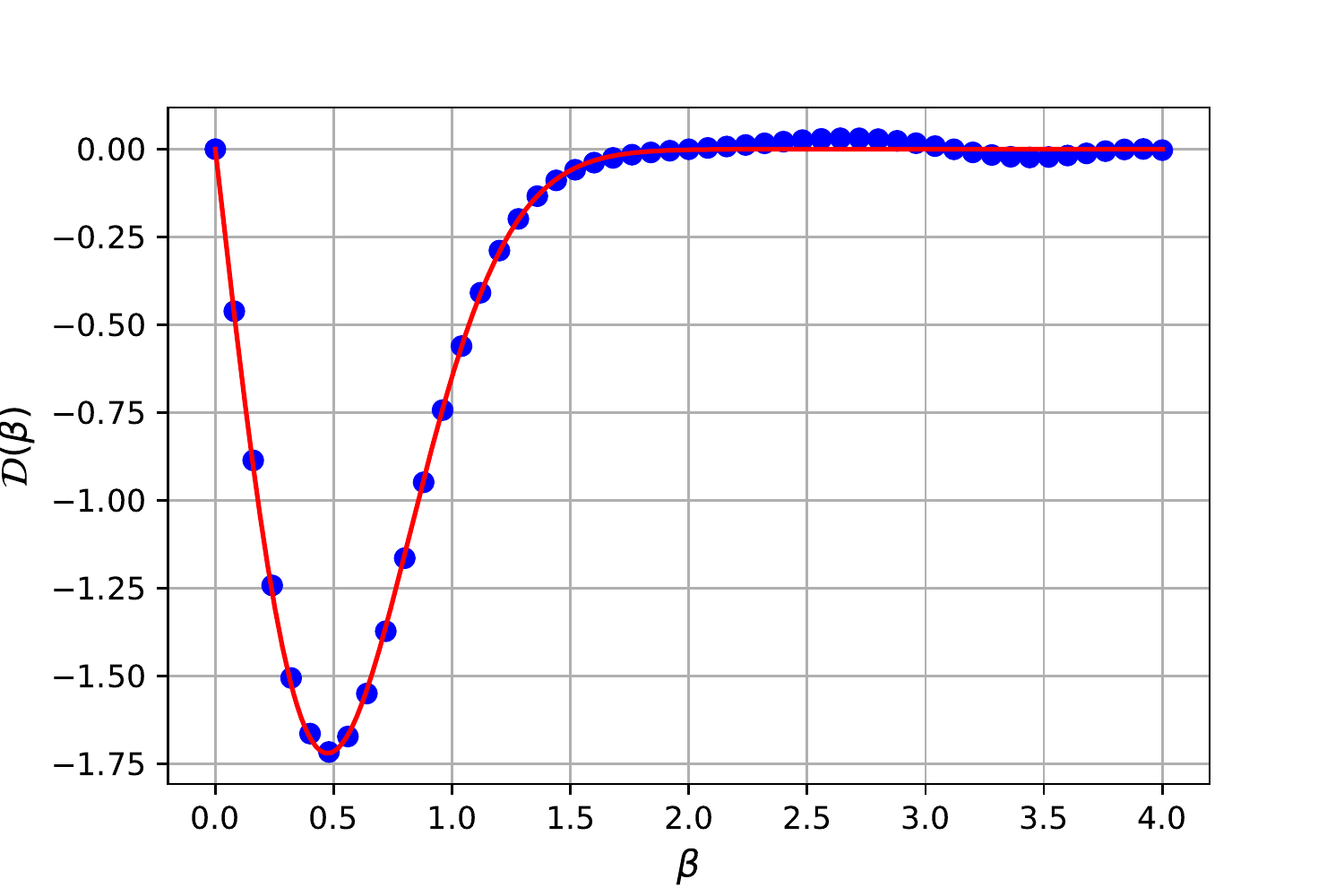} 
\includegraphics[scale=0.6]{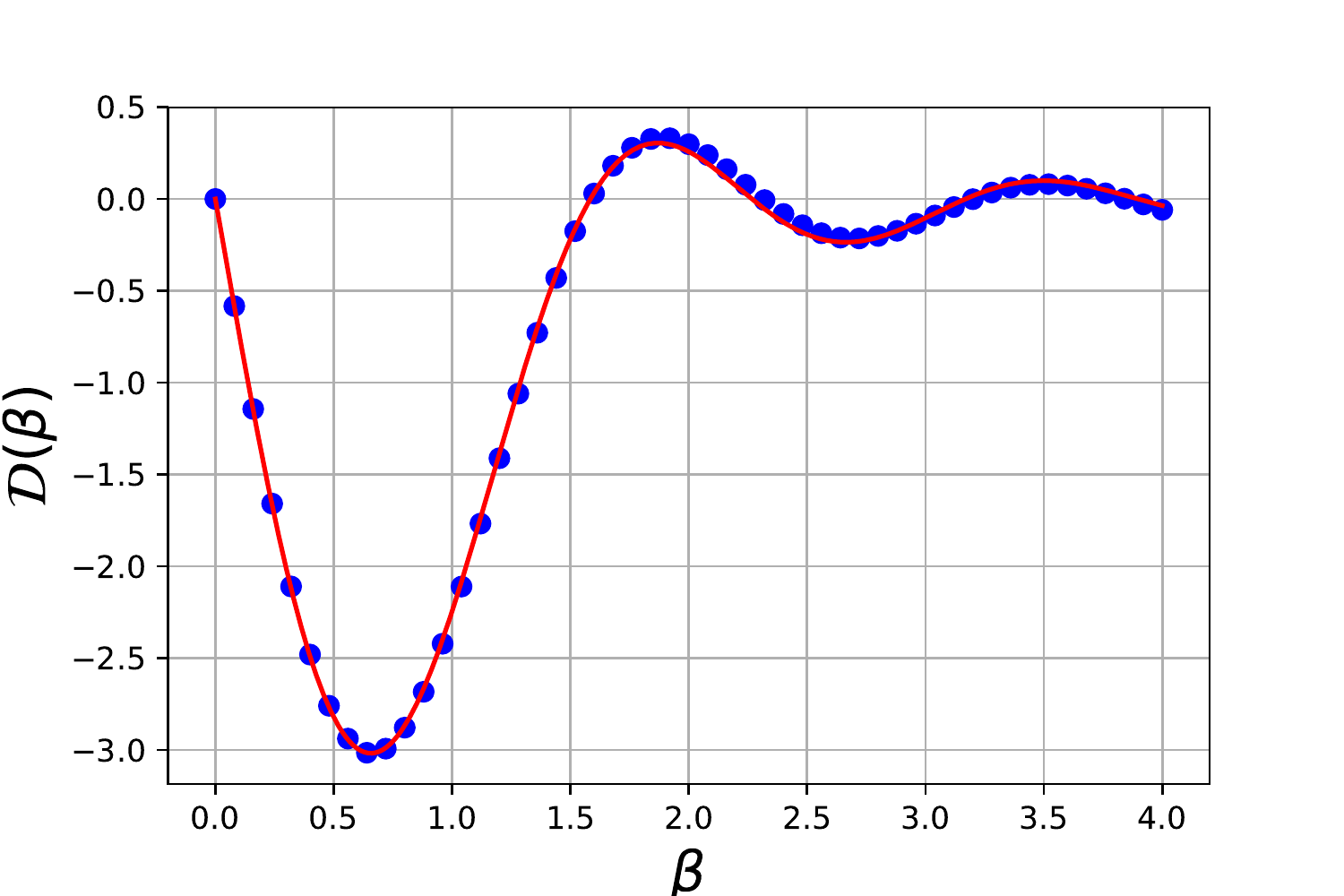} 
\caption{The aspect for the function ${\cal D}(\beta)$ given by (\ref{Dcall}) for 
a normal (top) and uniform (bottom) frequency  distributions $g(\omega)$ with null average. 
Top: A Barabasi-Albert network with 1000 nodes, average degree  $\langle d_k\rangle = 5.98$,
and a normal frequency distribution with  $\sigma_\omega = 3/2$.
Bottom: An Erd\"os-R\'enyi network with 1000 nodes, average  degree
 $\langle d_k\rangle = 7.39$, and a uniform frequency distribution in the interval $(-2,2)$.
 For both cases, 
the lines are the mean-field predictions, Eq. (\ref{Dcalln}) and Eq. (\ref{Dcallu}), 
and the blue circles   correspond to the respective numerical values calculated from (\ref{Dcall1}).   Notice that ${\cal D}'(0)$ is given by $-\cal L$,
 defined by (\ref{calL}).
}
\label{Fig3}
\end{figure}

Let us now focus on  the estimation of the synchronization threshold 
$\lambda_c$ from the mean-field approach. For sake of simplicity, let us consider the Kuramoto limit $\rho_k = \alpha$. 
Notice that  one can express
  \begin{equation}
  \label{Dcall1}
{\cal D}(\beta) = \frac{1}{N\sigma_\omega^2 } \sum_{k =1}^N \sum_{j =1}^N a_{kj} \omega_k \sin\beta(\omega_j-\omega_k)  
 \end{equation}
 in the mean-field approximation as
  \begin{eqnarray}
  \label{Dcall}
{\cal D}(\beta) &=&   \frac{\left\langle d_k \right\rangle}{\sigma_\omega^2}
\int d\omega g(\omega) \omega \int d\omega'g(\omega')\sin \beta(\omega'-\omega)\nonumber \\
&=&
\frac{\left\langle d_k \right\rangle}{2\sigma_\omega^2}
\frac{d}{d\beta} r^2(\beta),
  \end{eqnarray}
where $\langle d_k \rangle$ stands for the average degree of the network and $r(\beta)$, 
  the mean-field approximation for the order parameter (\ref{order_parameter}), is given by Eq. (\ref{rfixed}).
  Since equation (\ref{beta}) in the Kuramoto limit corresponds to
\begin{equation}
\label{beta1}
\dot\beta = 1 + \lambda {\cal D}(\beta),
\end{equation}
the value of $\lambda_c$ necessary to assure the existence of a zero 
for the right-handed side of
(\ref{beta1}) can be inferred from the minimum of ${\cal D}(\beta)$. 
One can easily determine the function $\cal D$ for normal and uniform
distributions, for instance. 
From the expressions (\ref{rn}), (\ref{ru}),  and (\ref{Dcall}), one has 
\begin{equation}
\label{Dcalln}
{\cal D}_{\rm n} = -\left\langle d_k \right\rangle\beta e^{-\beta^2\sigma_\omega^2}
\end{equation}
and
\begin{equation}
\label{Dcallu}
{\cal D}_{\rm u} = -\frac{\left\langle d_k \right\rangle}{ \beta \sigma_\omega^2 }
\left(\frac{\sin^2 \sqrt{3 }\beta\sigma_\omega }{3 \beta^2 \sigma_\omega^2} - \frac{\sin2\sqrt{3 }\beta\sigma_\omega}{2\sqrt{3 }\beta\sigma_\omega} \right) ,
\end{equation}
which respective aspects are depicted in Fig. \ref{Fig3} for some typical large random
networks. 
For both cases, we have a simple expression for the synchronization threshold,
\begin{equation}
\label{thresh}
\lambda_c = \gamma\frac{\sigma_\omega}{\langle d_k \rangle}.
\end{equation}
where $\gamma = \sqrt{2  e}$ for the normal distribution case.
For   uniform distributions,  one can   evaluate   $\gamma$ by noticing that  the global minimum of 
\begin{equation}
f(x) = \frac{d}{dx}\left( \frac{\sin x}{x}\right)^2.
\end{equation}
 can be easily determined numerically, and we will have 
finally  $\gamma\approx 2.14$.  Notice that $ \sqrt{2  e}\approx 2.33$ and, hence, the 
synchronization thresholds are rather close for both distributions. The key point, however,
is  the dependence of $\lambda_c$ on $\sigma_\omega$ and ${\langle d_k \rangle}$.

\subsection{The optimization algorithm}

We are now ready to define precisely 
 what we understand by the network synchronization capability and 
how to formulate an optimization scheme    in order to enhance it. Of course, we want to facilitate the appearance of fully synchronized regimes for the network, and this
can be achieved, for instance, demanding a smaller  threshold   value of    
 $\lambda_c$, which clearly would correspond to a network where a synchronization could occur more easily. Furthermore, we could also demand a better stability of the
fixed points $\left(\bar\rho_j,\bar\beta\right)$, which, on the other hand, would correspond to a more robust synchronized state.
The stability of the fixed point is determined by the eigenvalues of the Jacobian matrix 
$J=\frac{\partial (F_k,G)}{\partial(\rho_j,\beta)} $. 
We are mainly concerned with the fixed point near $\rho_k=\alpha$ and $\beta=0$ and, hence, we will approximate the Jacobian matrix at the fixed point $\left(\bar\rho_j,\bar\beta\right)$ by the 
Jacobian matrix at $\left(\alpha,0\right)$, 
\begin{equation}
\label{Jacobian}
J = \left[\begin{array}{c|c} 
 - (\lambda L + 2\alpha^2 \boldsymbol{I})  & 0 \\
\hline
-
\frac{2\Omega}{\alpha \langle \omega^2\rangle}\boldsymbol{\omega}^T & -\lambda {\cal L}
\end{array}\right].
\end{equation}
Since (\ref{Jacobian}) is a block matrix, its eigenvalues can be easily determined. Notice that any block matrix of the type (\ref{Jacobian})  can be decomposed as 
\begin{equation}
\label{decomp2}
  \left[\begin{array}{c|c} 
 A  & 0 \\
\hline
B& C  
\end{array}\right] = \left[\begin{array}{c|c} 
 A  & 0 \\
\hline
B& \boldsymbol{I} 
\end{array}\right]
\left[\begin{array}{c|c} 
 \boldsymbol{I}  & 0 \\
\hline
0 & C  
\end{array}\right],
\end{equation} 
where the consistent orders of the sub-hmatrices are implicitly assumed. 
The eigenvalues $\varpi$ of the Jacobian matrix $J$ corresponds to the roots of the
characteristic polynomial 
$
\det \left(J-\varpi \boldsymbol{I}\right) = 0,
$
which can be decomposed according to (\ref{decomp2}) as 
\begin{equation}
\det \left(J-\varpi \boldsymbol{I}\right) = 
  \left(\lambda {\cal L} + \varpi\right) \det \left((\lambda L + 2\alpha^2\boldsymbol{I})+\varpi \boldsymbol{I}\right) = 0.
\end{equation}
It is clear that the eigenvalues of the Jacobian matrix $J$  are $-\lambda {\cal L}$ and those ones of the $N\times N$ matrix  
$-(\lambda L + 2\alpha^2\boldsymbol{I})$. Moreover, 
since all of them are negative, the fixed point will indeed be always dynamically stable,
irrespective of the value of the rigid rotation $\Omega$. By demanding a larger value of  $\cal L$ given by (\ref{calL}), one simultaneously 
achieves both optimization conditions: a smaller effective threshold 
value   $\lambda_c$ and a more robust synchronized state. The maximization of ${\cal L}$   is exactly the optimization goal of the algorithm introduced in   \cite{optimal}, which we have just established to be also   valid   in  the present case of 
the SL model. 

Now, we can envisage a simple hill-climb rewiring optimization algorithm consisting, 
roughly,   in eliminating a random edge of the network  and substituting it with a new randomly chosen one. If the resulting network is still connected and has a larger value of $\cal L$, the modification is accepted and the procedure is repeated. This algorithm typically produces
networks with far  better synchronization capabilities, and its usage requires quite modest computational resources, even for large networks, since (\ref{calL}) is a simple quadratic function.   For our purposes here, we will use $\cal L$ as a quantifier for the network synchronization capability, the larger value of $\cal L$, the better synchronization properties 
of the network. This conclusion is totally compatible with the numerical works done before
for the Kuramoto model \cite{louzada2012,li2013,zhang2013} and also
for the so-called Kuramoto model with inertia \cite{pinto2014}, which is particularly relevant 
to the study of power lines\cite{filatrella2008,motter2013}. We have confirmed here,  by exhaustive numerical simulations,
the validity of this conclusion also for the SL case. 

\begin{figure}[t]
\includegraphics[scale=0.6]{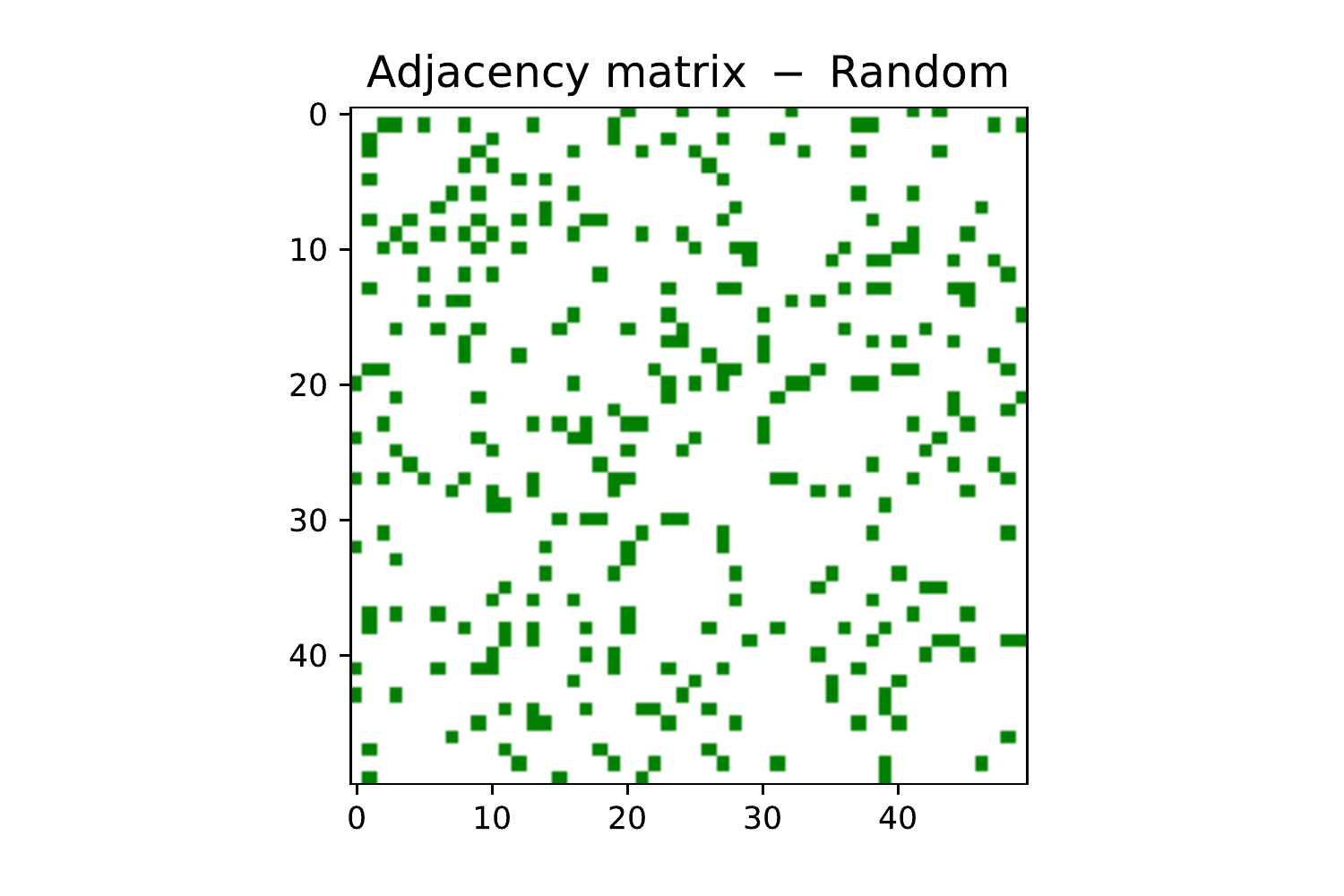} 
\includegraphics[scale=0.6]{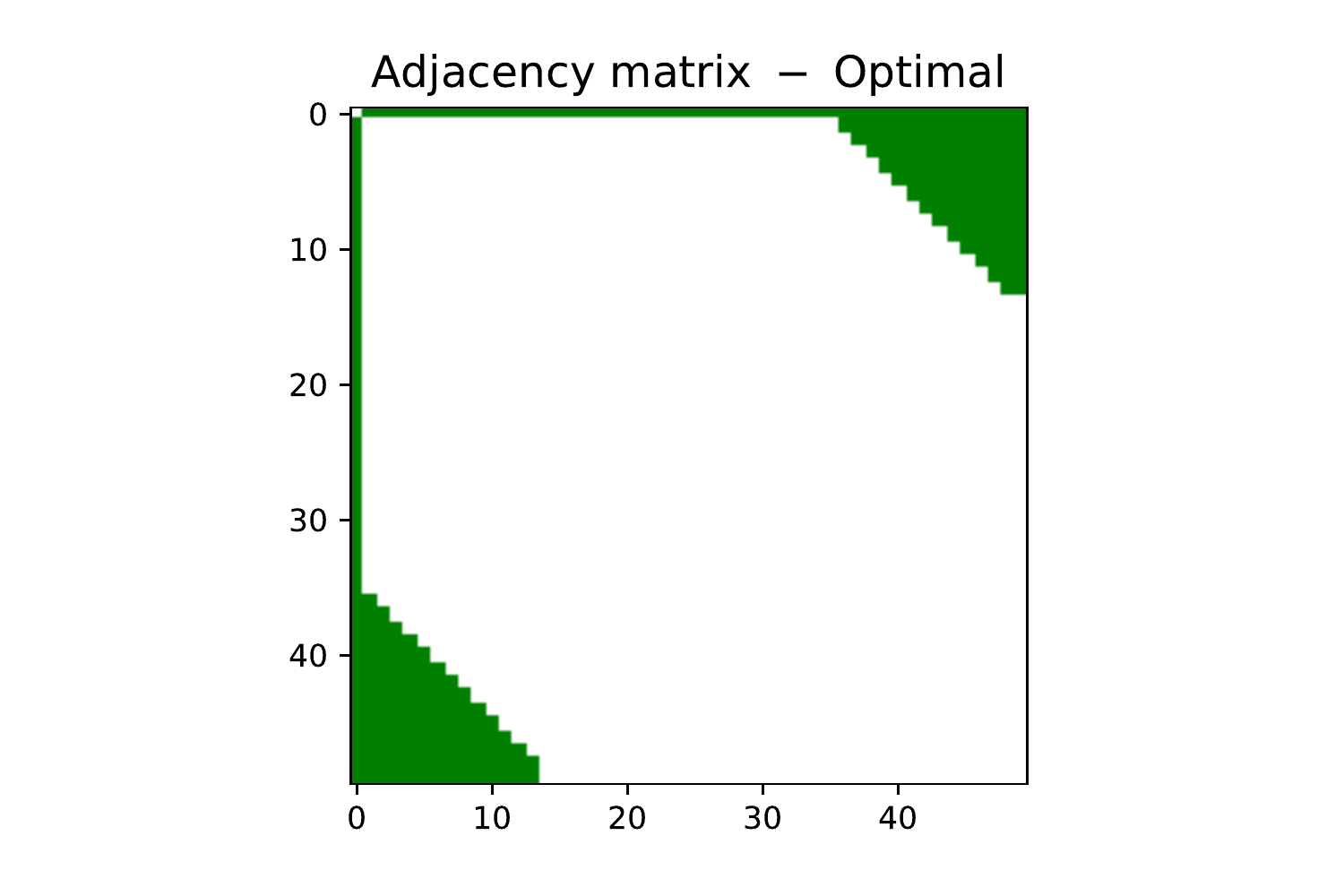} 
\caption{Top: nonzero entries of the adjacency matrix for a random Erd\"os-R\'enyi network with 50 nodes and 153 edges ($\langle d_k \rangle = 6.12$). The nodes are sorted and numbered according to the corresponding values of $\omega_k$, which is assumed to be a random variable. The edges are randomly scattered in
the matrix, implying the vanishing of the last term in the Eq. (\ref{decomp1}). 
Bottom: the corresponding optimal rewiring for the same network. The edges are now concentrated in the corners of the matrix, and we can estimate (\ref{decomp1}) by restricting the sum to the area occupied by the edges in the matrix. 
}
\label{Fig4}
\end{figure}

\begin{figure}[t]
\includegraphics[scale=0.6]{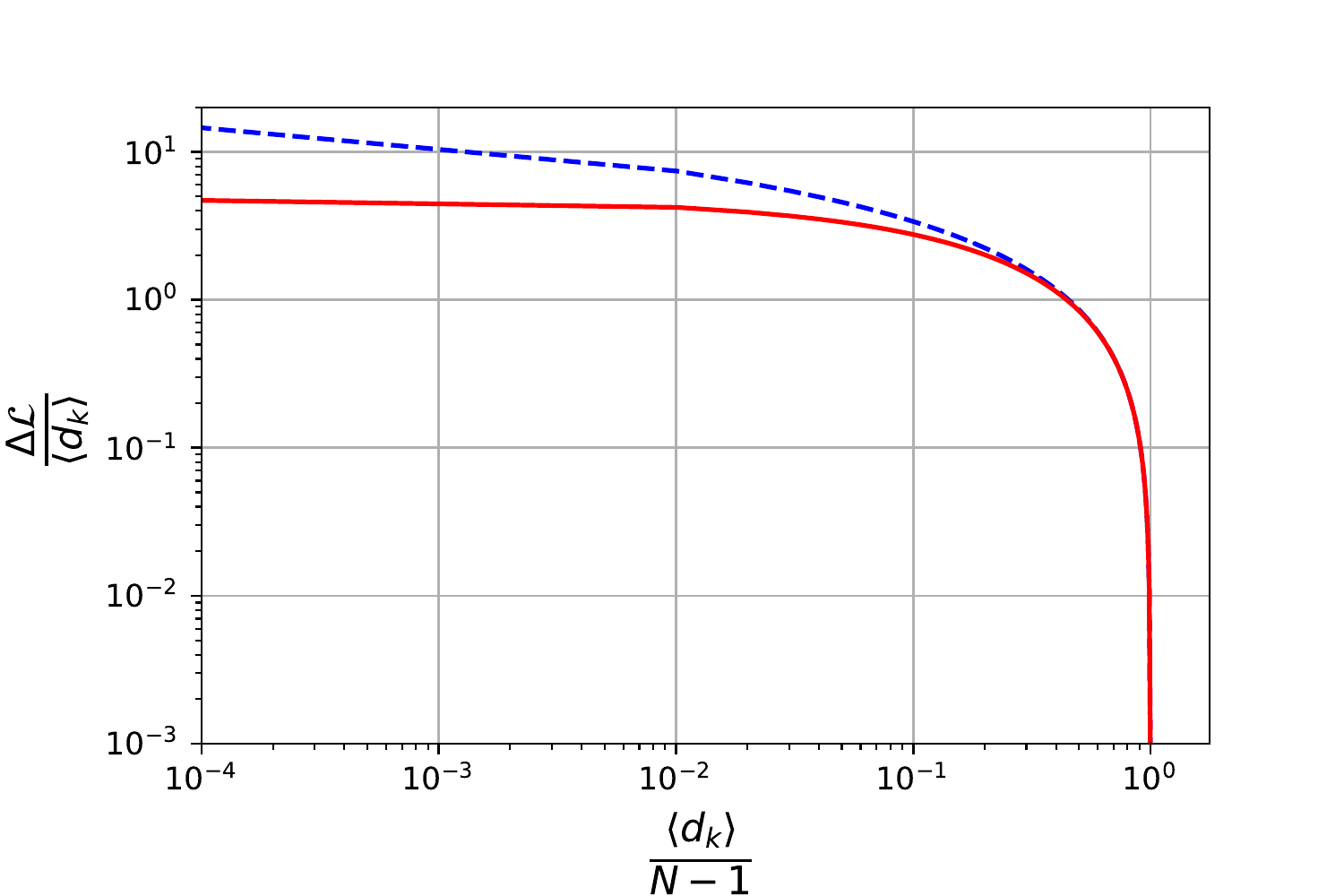} 
\caption{The maximum gain (\ref{gain}) in the network synchronization capability achieved
by using the optimization algorithm. The (blue) dashed and the (red) solid lines correspond, respectively, to the normal and uniform distribution cases.   The algorithm is typically much more efficient for sparse networks. For instance, for the networks of Fig. \ref{Fig3}, one has 
$\frac{\langle d_k \rangle}{ N-1 } \approx 0.006\sim 0.008$, and ${\cal L}$  could be enlarged by a factor of 5 and 8, respectively, by using the algorithm for the uniform and normal
distribution cases}
\label{Fig5}
\end{figure}

We can improve considerably the performance of our algorithm by exploring some
heuristics.  For instance,  any Laplacian matrix can be decomposed as
the sum of elementary   matrices for the edges
\begin{equation}
\label{decomp}
L = \sum_{e(i,j)}L_{(i,j)},
\end{equation}
where the sum is to be performed over all the edges $e(i,j)$ in the network, and $L_{(i,j)}$ is the elementary Laplacian matrix corresponding to the sole edge connecting the nodes $i$ and $j$. From
the decomposition (\ref{decomp}), we have that (\ref{calL}) can be written as
\begin{eqnarray}
\label{decomp1}
{\cal L} = \sum_{e(i,j)}\frac{\boldsymbol{\omega}^TL_{(i,j)}\boldsymbol{\omega} }{\boldsymbol{\omega}^T\boldsymbol{\omega}} &=&  \sum_{e(i,j)} 
\frac{(\omega_i-\omega_j)^2 }{\boldsymbol{\omega}^T\boldsymbol{\omega}} \\
 &=& \sum_{k=1}^{N} \frac{d_k\omega_k^2}{\boldsymbol{\omega}^T\boldsymbol{\omega}}-2\sum_{e(i,j)} 
\frac{\omega_i\omega_j }{\boldsymbol{\omega}^T\boldsymbol{\omega}},\nonumber
\end{eqnarray}
from where one can immediately recognize if a certain rewiring step in our algorithm will be successful or not. Besides of keeping the connectedness of the network, the value of 
$|\omega_i-\omega_j|$ for the new link must be larger than   the original one.
For a random network with random frequencies, one can also estimate ${\cal L}$   from a mean field approximation.
For symmetric frequencies distribution $g(\omega)$ with null average, 
the last term in (\ref{decomp1}) vanishes, leading to 
${\cal L} \approx \langle d_k \rangle$, which we have indeed corroborated in our numerical simulations. Also  from a mean field approximation, 
we can
estimate  ${\cal L}_{\rm max}$, the largest possible value for ${\cal L}$ obtained by
applying  our algorithm to a random network. It will correspond to the connected network where the weakest edges (smallest values of  $|\omega_i-\omega_j|$) where
substituted with the strongest ones (largest values of  $|\omega_i-\omega_j|$). In this case, the last term in (\ref{decomp1}) does not vanish anymore   since the edges are no more randomly scattered in the adjacency matrix, see  Fig. \ref{Fig4}.  
Let us suppose we have a
random network with average degree  $\langle d_k \rangle$, where
 its $N$ nodes have been sorted and numbered accordingly to the respective values of
 the random variable $\omega_k$. The number of edges will be
$N_{\rm e} = N \langle d_k \rangle/2$ and we can estimate ${\cal L}_{\rm max}$ 
for a normal distribution of frequencies $g(\omega)$
as (see Fig. \ref{Fig4})
\begin{equation}
{\cal L}_{\rm max}^{\rm (n)}  \approx  \frac{2N}{\sigma^2_\omega}\int_{-\infty}^\infty du\,
g(u) \int_{\frac{\omega_*}{\sqrt{2}}}^\infty dv\, g(v) v^2 ,
\end{equation}
where $\omega_*$ is such that
\begin{equation}
 \int_{-\infty}^\infty du\,
g(u) \int_{\frac{\omega_*}{\sqrt{2}}}^\infty dv\, g(v)  = \frac{\langle d_k \rangle}{2(N-1)},
\end{equation}
which leads to
\begin{equation}
\frac{1}{N}{\cal L}_{\rm max}^{\rm (n)} \approx  \frac{\langle d_k \rangle}{N-1}  + \frac{2\kappa e^{-\kappa^2}}{\sqrt{\pi}},
\end{equation}
with
\begin{equation}
{\rm erfc}(\kappa) = \frac{\langle d_k \rangle}{N-1}.
\end{equation}
In an analogous way, one can obtain for the case of a uniform frequency distribution $g(\omega)$
\begin{equation}
\frac{1}{N}{\cal L}_{\rm max}^{\rm (u)} =  \frac{\langle d_k \rangle}{ N-1 } \left(
3\frac{ \langle d_k \rangle}{ N-1 } -8 \sqrt{\frac{ \langle d_k \rangle}{ N-1 }}
+6
 \right).
\end{equation}
By construction, our algorithm preserves the total number of edges and, consequently,
also the average degree
$\langle d_k \rangle$. For a network with $N$ notes, the maximum possible value for $\langle d_k \rangle$ is $N-1$, which corresponds to the {\em all-to-all} connection topology. In this case, there is no room for optimization and ${\cal L}_{\rm max} \approx \langle d_k \rangle$ for large $N$. Fig. \ref{Fig5} depicts the maximum gain in the network synchronization capability, 
defined as
\begin{equation}
\label{gain}
 \Delta {\cal L} = {\cal L}_{\rm max} - \langle d_k \rangle ,
\end{equation}
obtained by employing our algorithm on a large random network with average degree
$\langle d_k \rangle$. The algorithm is typically much more efficient for sparse networks, {\em i.e.}, for networks with $ \frac{\langle d_k \rangle}{ N-1 } \ll 1$.

\section{Interlayer symmetries and synchronization} 

Let us now consider the synchronization capability of multilayer networks as those ones
depicted in Fig. \ref{Fig1} in the light of the results of last Section. We are
mainly interested here in   multilayer networks built from identical layers, which are supposed to have $N$ nodes each and   Laplacian matrix  $L^{(0)}$.
The Laplacian matrix of a generic 
$k$-layer  networks built from such identical   layers   will be given by the following $kN\times kN$ symmetric matrix
\begin{equation}
\label{klayer}
L=\left[\begin{array}{c|c|c|c}\displaystyle
L^{(0)} + D_1   & - {C}_{12} & \cdots & -C_{1k} \\
\hline
-C_{21} & L^{(0)} + D_2 & \cdots & -C_{2k} \\
\hline
\vdots & \vdots & \ddots & \vdots \\
\hline
-{C}_{k1}& \cdots & \cdots & L^{(0)} + D_k 
\end{array}\right],
\end{equation} 
where $C_{ij}$ is an integer matrix with entries 0 or 1,
corresponding to the edges connecting the layer $i$ to the layer $j$. Let us call it the
interlayer connection matrix. Since we are considering only undirected networks,  we have 
by construction  $C_{ij} = C_{ji}^T$.
The matrix $D_i$ is a diagonal matrix  which entries stand for the sum of the corresponding lines of the interlayer  connection matrices $C_{ij}$, with $1 \le j \le k$, $i\ne j$. 
It is clear that the {  diagonal} connection pattern discussed in Fig. \ref{Fig1} does correspond to the case where all the interlayer connection matrices $C_{ij}$ are diagonal.

Roughly, a symmetry of a network, also called a graph automorphism in the mathematical
literature,  is a permutation of some of its nodes which preserves the network connection structure. This is equivalent to state that 
a permutation matrix $P$  will correspond to a network symmetry if and only if $[L,P]=0$, where $L$ is the network Laplacian matrix. The full set of symmetries  of a network defines   the so-called automorphism group, which determination for generic cases 
 is typically a computationally  complex problem. 
Nevertheless,  some very efficient algorithms for finding  graph automorphisms in
concrete situations are available as, for instance, the {\tt nauty} package\cite{Nauty}. 
For our purposes here, we will consider symmetries of the SL model (\ref{SL}), {\em i.e},
the automorphisms of the underlying network which also preserves the oscillator frequencies.
In other words, finding the symmetries of a SL system is equivalent to the so-called 
  colored graph isomorphisms problem, see \cite{Nauty} for further references. Thus, 
 a permutation $P$ of nodes will be a symmetry of (\ref{SL}) if, besides commuting with the network Laplacian matrix, it also obeys $\boldsymbol{\omega} = P \boldsymbol{\omega}$,
where     $\boldsymbol{\omega}$ for a multilayer network with $k$ identical layers 
is the $kN$-dimensional vector formed by $k$ copies the oscillator natural frequencies 
of each layer. 
  In this case, the node permutations corresponding to the matrix
 $P$ will effectively lead to dynamically equivalent SL systems.
 (We have also the possibility $\boldsymbol{\omega} = -P \boldsymbol{\omega}$, we will return to this point in the last section.)  
  Since we assume the natural
 frequencies $\omega_k$ to be random variables, symmetries are typically very rare in our context. 
  However, for some cases it is quite easy, or
even natural, to have some permutation symmetries. These are precisely
the case of multilayer networks of oscillators with identical
layers, which are the focus of the present work. Such a kind of
structured network has been intensively investigated recently, and there are a myriad of possible applications in many
areas, see \cite{multi1,multi2} for recent comprehensive reviews. 
The existence of entire-layer permutation symmetries is the main reason why such specific type of structured networks are the relevant ones for our analysis.

\begin{figure}[ht]
\includegraphics[scale=0.5]{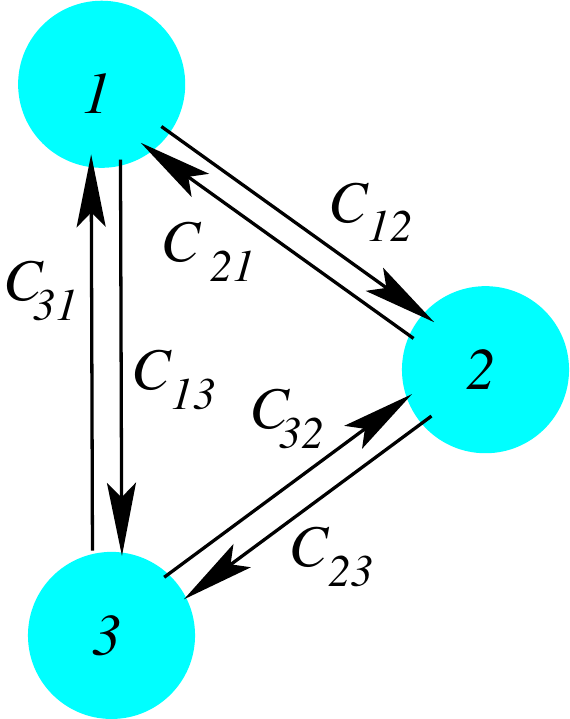} 
\caption{Block representation of the trilayer network of Fig. \ref{Fig1}. The diagonal
interlayer connection pattern corresponds to the case where the connection matrices $C_{ij}$ are diagonal, and hence the block network is effectively undirected, since one has always $C_{ji} = C_{ij}^T = C_{ij}$
 in this case.
  For the non-diagonal connection 
pattern, we have, in general, $C_{ji} = C_{ij}^T\ne C_{ij}$ and, in this case, the block network corresponds effectively to a directed    network.  In both cases, the
effective block representation will be in general a weighted (edge-labeled) network. The unweighted case corresponds to the situation where all the connection matrices are equal. }
\label{Fig6}
\end{figure}

Let us now focus on the connections between two arbitrary layers $i$ and $j$. The equivalent of the elementary Laplacian matrix for these two layers has the following form
\begin{equation}
L_{(i,j)}=\left[\begin{array}{c|c|c|c|c}\displaystyle
\cdots   &  {\cdots} & \cdots &  {\cdots}   & \cdots\\
\hline
\  \cdots  & L^{(0)} + D_i  & \cdots & - {C}_{ij}&\cdots \\
\hline
\vdots & \vdots & \ddots & \vdots&\cdots \\
\hline
\cdots& - {C}_{ij}^T & \cdots & L^{(0)} + D_j  &\cdots \\
\hline
\cdots   & \cdots & \cdots & \cdots  & \cdots
\end{array}\right],
\end{equation} 
where only the non-zeros components are showed. Let $P_{(i,j)}$ be 
the matrix corresponding to the permutation of the entire layers $i$ and $j$.  
It is easy to check that $\left[L_{(i,j)},P_{(i,j)}\right]=0$ does require $C_{ij}=C_{ij}^T$.
In other words, the elementary Laplacian matrix  of the layers $i$ and $j$  will be invariant under the permutation of the entire layers  $i$ and $j$ if and only if the corresponding connection matrix were symmetric. Now, one can grasp the real peculiarity of the diagonal interlayer connection: it always leads to the permutation symmetry of the corresponding elementary Laplacian matrix. A generic connection pattern, of course, will correspond to a non-symmetric $C_{ij}$ and, in this case, the permutation symmetry is absent in general. 
Entire layer permutations of the  full  Laplacian matrix (\ref{klayer}) can be 
 seen as permutations in a directed and weighted (edge-labeled) network with $k$ nodes, each node corresponding to a layer in the original network, see Fig. \ref{Fig6}, where the  networks of Fig. \ref{Fig1} are now depicted as   block networks,   with each block corresponding to an entire layer.
The necessary conditions for a permutation be a symmetry in directed weighted networks are of course much more restrictive than in the undirected unweighted case.  For instance, the permutation $P_{(i,j)}$ discussed above will be a symmetry of the elementary Laplacian 
matrix $L_{(i,j)}$ only if the edge connecting the layers $i$ and $j$ were effectively 
undirected, {\em i.e.}, $C_{ij} = C_{ij}^T = C_{ji}$. Nevertheless, this is not enough to assure that $P_{(i,j)}$  be a symmetry of the entire network. For instance, a permutation of the levels 1 and 2 in the block-network of Fig \ref{Fig6} would be a symmetry only if, besides of $C_{12} = C_{12}^T = C_{21}$, we have
$C_{31}=C_{32}$ and $C_{13}=C_{23}$. 

\subsection{The results}

We are now ready to state  our main results. We have considered   multilayer
networks with the Laplacian matrix given by (\ref{klayer}). 
The identical layers are generated by using the NetworkX \cite{NetworkX}
package  for python, which allow us to build many types of random networks with prescribed
topology and statistical properties. The connections between the layers are also randomly
chosen with different statistical properties, mimicking  in this way the situations where
the inlayer and interlayer connection patterns evolve and are selected differently.   
The natural frequencies are draw according to the prescribed distributions, and the  corresponding governing equations (\ref{SLR1}) and (\ref{SLR2}) are then solved with the help of the 
 SciPy \cite{SciPy} package for python. 

Since all  the layers in our network are   identical, if the interlayer connection matrices $C_{ij}$ were diagonal, the corresponding interlayer connections will not contribute to the parameter ${\cal L}$ given by (\ref{decomp1}), which is precisely the quantity to be maximized in our optimization procedure. 
For these cases,   one has
${\cal L} =   {\cal L}_0$, where ${\cal L}_0$ is the contribution to  (\ref{decomp1}) from  each isolated layer and, hence,
 the synchronization capability of the whole network is effectively the same of each isolated layer.  In this case, one is not taking any advantage of the multilayer structure and, in this sense,
this is the {\em worst} possible configuration for synchronization in these networks.
 However, starting with a  multilayer
network  with diagonal interlayer connections, one may apply our algorithm only to the
interlayer edges between two given layers and, in consequence, one will enhance considerably the network synchronization capability (increasing the value of ${\cal L}$, accordingly to the results of the Section II), whereas keeping the layers and the number of connections between them unchanged. This can be considered a minimal change in the network topology, but with a considerable impact on its synchronization capability. 
We will consider as optimal the network obtained from a full rewiring of the interlayer connections. Of course, one should expect a gradual improvement of the network synchronization capability by considering larger and larger rewiring ratio, but a more quantitative description of this process is still lacking.

\begin{figure}[t]
\includegraphics[scale=0.6]{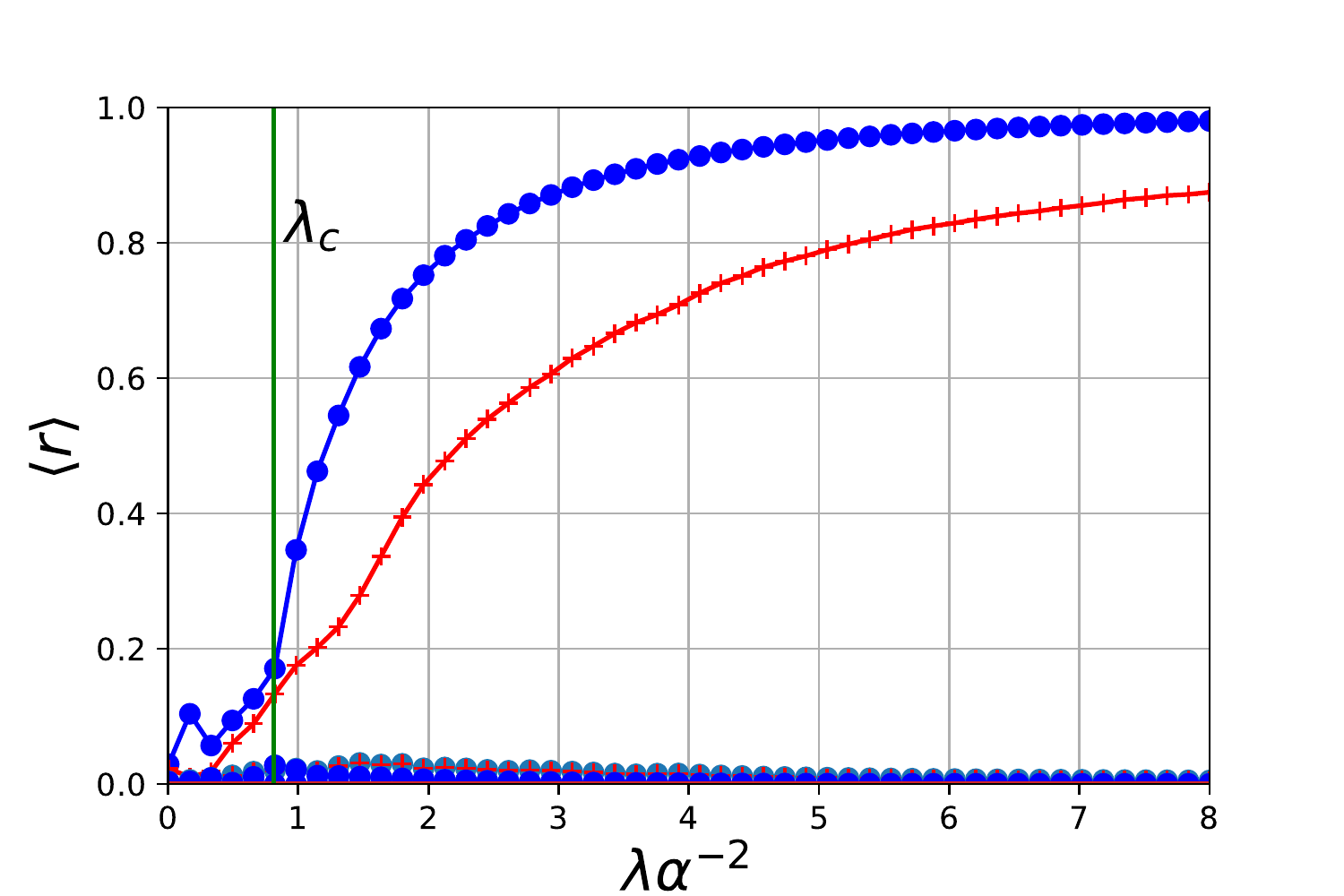} 
\caption{Synchronization diagram for a
bilayer random network with 800 nodes disposed in two identical 
 Barabasi-Albert layers, with 352 interlayer connections. (Interlayer
connection probability $p=0.9$, see the text for details.)
The oscillators native frequencies were drawn from a normal distribution with $\sigma\alpha^{-2} =1$.
The entire network has average degree $\langle d_k\rangle = 2.88$, and the
minimal and maximal degrees are, respectively, 1 and 24.  The synchronization
threshold 
  $\lambda_c\alpha^{-2}=0.81$, given by (\ref{thresh}),  is depicted as a vertical line. 
The red crosses corresponds
to the original network with diagonal interlayer connections, for which ${\cal L}=2.11$, while the blue circles corresponds to a network with exactly the same layers, but with optimal interlayer connections, for which    ${\cal L}=11.48$. 
The optimal network has, by construction, the same average degree, but the
minimal and maximal degrees are now, respectively, 1 and 71.
 The curve and points at the bottom of the graphics correspond to the standard deviation associated with the average (\ref{aver}), which assures that the average $\langle r\rangle$ is being indeed evaluated in the stationary regime. It is clear from the diagram that the optimal non-diagonal interlayer connections has enhanced considerably the synchronization capability of the original network. 
}
\label{Fig7} 
\end{figure}

We have performed exhaustive numerical simulations which have corroborated our main conclusions. Our analysis starts with a multilayer random network with diagonal interlayer connection, for which  we construct a synchronization diagram $\langle r\rangle\times \lambda$ (see Fig. \ref{Fig7} for instance), 
where
  $\langle r\rangle $ corresponds to the average of the   order parameter (\ref{order_parameter}) evaluated on some time interval $\Delta t$   
 \begin{equation}
 \label{aver}
 \langle r\rangle = \frac{1}{\Delta t} \int_{t_0}^{t_0+\Delta t} r(t)\, dt ,
\end{equation}  
along the corresponding numerically integrated  solutions of (\ref{SL}), starting with
random initial conditions. The parameter  
 $t_0$ is meant to be chosen in order to ensure the dissipation of any transient regime, and this can be monitored, for instance, from the standard deviation of $r$ on the integration interval. 
It is more convenient to work with the dimensionless evolution parameter $\tau = \alpha^2 t$, which of course corresponds to set $\alpha=1$, and $\lambda = \lambda\alpha^{-2}$ and
$\omega_k = \omega_k\alpha^{-2}$ in (\ref{SL}). It is clear that in this case we are invoking
$\alpha^{-2}$ as the intrinsic time scale for our problem. 
For a fixed multilayer random network, one evaluates   $\langle r\rangle $ for different
values of $\lambda$
and depict the final diagram as a graphics  $\langle r\rangle\times \lambda$.
We then apply our algorithm for the interlayer connections, generating a new network, and
repeat the same steps. By comparing the diagrams  $\langle r\rangle\times \lambda$ of
both networks, one can determine firmly which one has the best synchronization capabilities,
and it always is the optimal non-diagonal case.

\begin{figure}[t]
\includegraphics[scale=0.6]{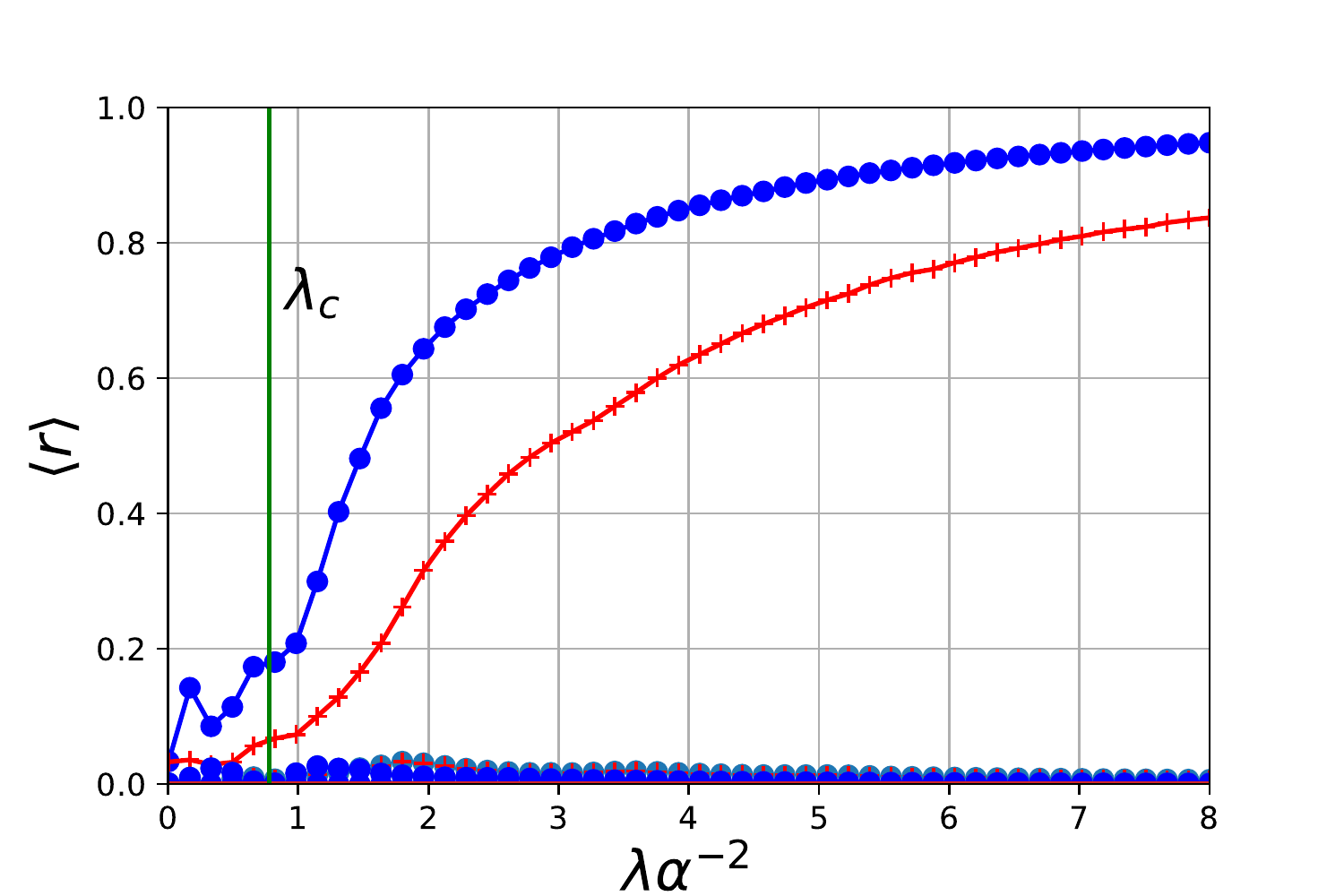} 
\caption{Synchronization diagram, with the same conventions of Fig. \ref{Fig7}, 
 for a random network with 1200 nodes disposed in three identical  Barabasi-Albert  layers, with 192, 216, and 188 interlayer connections, corresponding to $p=0.5$.
The trilayer network has average degree $\langle d_k\rangle = 2.99$, synchronization
threshold 
  $\lambda_c\alpha^{-2}=0.78$, and  the
minimal and maximal degrees are, respectively, 1 and 55. 
The original network with diagonal interlayer connections (red crosses) has ${\cal L}=1.47$,
 while  ${\cal L}=13.11$ for the optimal network (blue circles), for which the
minimal and maximal degrees  are, respectively, 1 and 74.
}
\label{Fig8}
\end{figure}

\begin{figure}[t]
\includegraphics[scale=0.6]{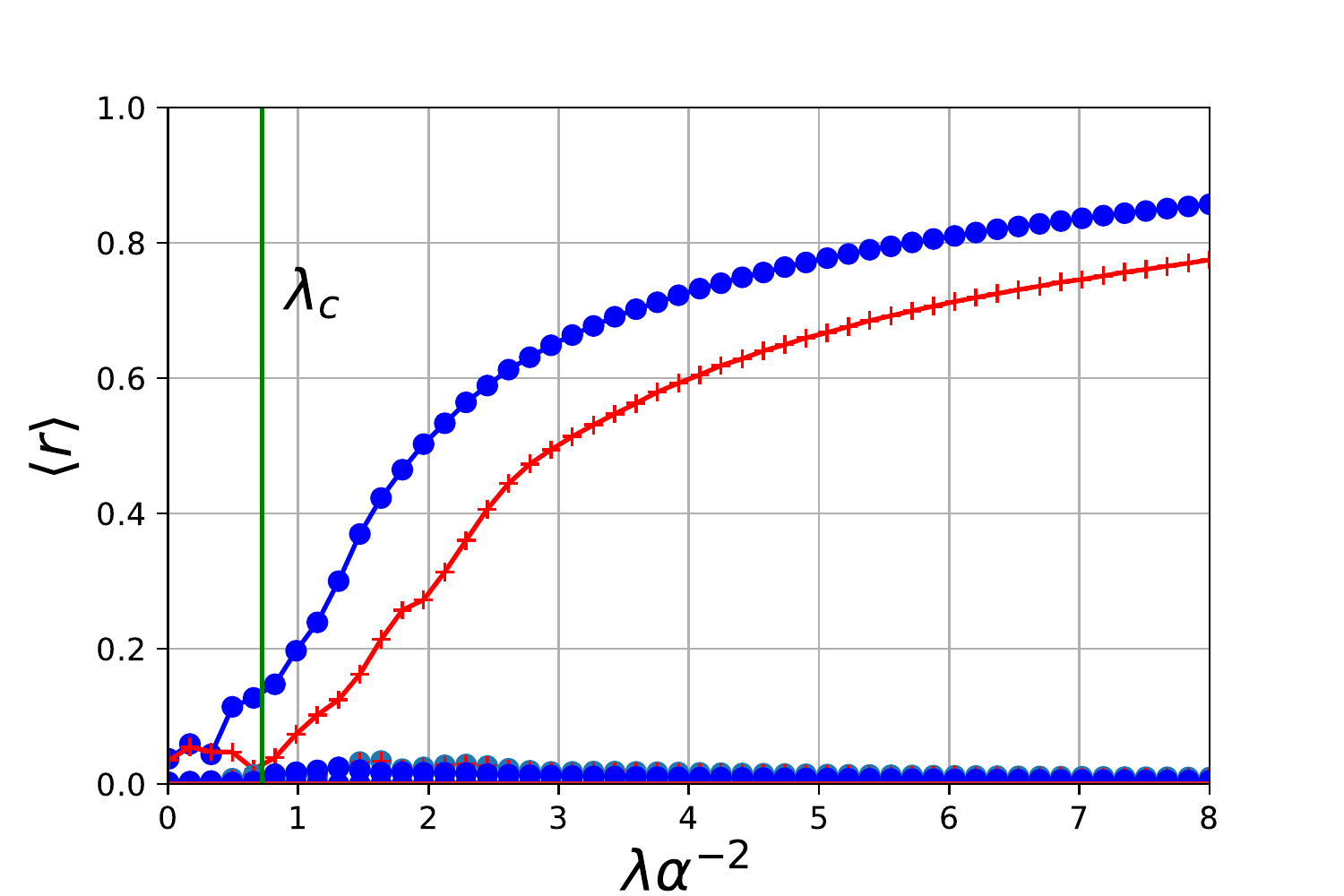} 
\caption{Synchronization diagram, with the same conventions of Fig. \ref{Fig7}, 
 for a random network with 2400 nodes disposed in six identical  Barabasi-Albert layers, with 92, 104, 86, 104, 89, 106, 115, 83, 106, 108, 77, 100, 116, 85, 100  interlayer connections, corresponding to $p=0.25$.
The six-layer network has average degree $\langle d_k\rangle = 3.22$, synchronization
threshold 
  $\lambda_c\alpha^{-2}=0.72$, and the
minimal and maximal degrees are, respectively, 1 and 42. 
The original network with diagonal interlayer connections (red crosses) has ${\cal L}=1.58$,
 while  ${\cal L}=16.05$ for the optimal network (blue circles), for which the
minimal and maximal degrees  are, respectively, 1 and 58.
}
\label{Fig9}
\end{figure}

According  to the discussions of the last Section, one should expect better results for sparse networks. Figs \ref{Fig7}, \ref{Fig8}, and \ref{Fig9}   
depict some typical cases built on  Barabasi-Albert
random networks \cite{barabasi1999} corresponding, respectively, to a bilayer, a trilayer, and a six-layer networks. 
For all cases, one starts with a 400-nodes  Barabasi-Albert  network corresponding to the layers and, then, a multilayer network with diagonal interlayer connection is constructed. The interlayer connections are also random and the probability of two equivalent nodes in different layers be connected is given by $p$. We adopt for the    bilayer,   trilayer, and   six-layer networks, respectively, $p= 0.9$, $0.5$, and $0.25$.
For all case, $\alpha^2t_0 = 4.25$ was enough to ensure the onset of the stationary regime  and $\alpha^2\Delta t = 0.75$ was used to evaluate $\langle r\rangle$  accordingly to (\ref{aver}). It is quite clear that the asymmetrical connection patterns 
systematically 
give rise to multilayer networks with far better synchronization capabilities.

\section{Final remarks}  

In this paper, we  have considered  the synchronization problem for Stuart-Landau oscillators 
in random 
   multilayer networks with 
identical layers.  We have shown that the breaking
of interlayer connection symmetries leads typically to an appreciable 
enhancement of the network synchronization capabilities, in the
same spirit of Motter-Nishikawa phenomenon of 
asymmetry-induced synchronization (AISync) \cite{NishMotter}. 
The dynamics of multilayer networks can be
conveniently  formulated in terms
of effective block-networks, where  a node stands for an entire layer 
of the original multilayer network\cite{multi1,multi2}. 
The presence of asymmetries 
in the interlayer connection patterns effectively transforms the problem  into a
directed   block-network (see Fig. \ref{Fig6}) which has typically 
  far better synchronization capabilities than   similar networks
with symmetric interlayer connection pattern,
which, on the other hand, are effectively described by undirected block-networks. 
  Moreover, the most regular connection pattern, corresponding to diagonal interlayer connection matrices (see Fig. \ref{Fig1}), is  always   the worst possible    configuration for synchronization in
these networks. In this sense, our results  
 corroborate the  main
conclusions of \cite{ZhangNishMotter}, namely that asymmetry-induced 
synchronization  should be a rather generic property of structured networks. In fact,
for unstructured networks with random frequency oscillators, symmetries should be extremely rare
since, despite of preserving the connection topology, a node permutation must also preserve the oscillator native frequencies in order to be a genuine symmetry for the SL system. 
The problem of finding the symmetries of a SL system is equivalent to   the   so-called colored graph isomorphisms problem, which are indeed much more restrictive than the usual  graph isomorphisms   \cite{Nauty}. In particular, if all the oscillators have different
native frequencies, which is a quite common situation if the   frequencies are assumed to be random real numbers,  there will be no permutation symmetry for the system.  In this sense, one should not expect  in general an
anti-correlation between symmetries and synchronization for unstructured random networks
of oscillators mainly because one should not expect any symmetry for them.

Our multilayer random networks with identical layers are explicitly built
in order to assure that some permutations of entire layers could in principle be genuine symmetries
for the SL system. Since all the
layers are identical, we have in this case $\boldsymbol{\omega} =  P \boldsymbol{\omega}$,
where $P$ is the pertinent entire-layer permutation matrix   and $\boldsymbol{\omega}$ stands for the oscillator native frequencies vector for the whole network or, in other words, the oscillator native frequencies for the whole network are even under the action of $P$. However, a situation where a certain permutation $P$ preservers the Laplacian matrix, {\em i.e.} $[L,P]=0$, and 
the native frequencies are odd under its action, {\em i.e.}
$\boldsymbol{\omega} =  - P \boldsymbol{\omega}$, can be also considered a symmetry for the SL system (\ref{SL}), since under $P$ we will have a system such that $z_k\to \bar z_k$, which of course has the same synchronization properties. Such a system cannot have identical layers, in fact it must have layers with the same topology, but with  reversed oscillator native frequencies. For these systems, the diagonal connection pattern is not anymore the worst possible case for synchronization, since the contribution from the interlayer edges to 
${\cal L}$ does not vanish as for the even case, see (\ref{decomp1}). This kind of systems is quite interesting and would deserve a deeper analysis. 

We would like also to recall  that 
we opt in our analysis to keep the layers unchanged and to rewire the edges between then
in order to optimize the synchronization. 
This could be considered a minimal change in the network topology, but typically with a considerable impact on its synchronization capability. 
This is not, however, the only optimization scheme one can envisage for the network. One could, for instance, to keep the edges unaltered and then to redistribute the oscillators native frequencies over the network. Let us suppose one keeps the layers identical in order to guarantee some possible permutation symmetries for the system. This problem can be formulated as follows: for a given Laplacian matrix $L$ in (\ref{calL}), which reordering of the vector $\boldsymbol{\omega}$ would lead to a maximal 
${\cal L}$? It is clear from (\ref{calL}) that the maximum possible value for ${\cal L}$ corresponds to the situation where $\boldsymbol{\omega}$ is proportional to the eigenvector of $L$ with largest eigenvalue, say $\boldsymbol{\nu}_{\rm max}$, which, incidentally, 
is an optimal synchronization condition that 
 has been already discovered
numerically and by means of more intricate methods \cite{brede2008a,buzna2009,kelly2011,skardal2014}. 
 Hence, an optimal reordering of $\boldsymbol{\omega}$ would be that one which minimizes $\Gamma=\left|\boldsymbol{\omega}-\boldsymbol{\nu}_{\rm max}\right|$. However, this problem is much more complex computationally than the rewiring problem we have discussed in this paper. This issue is now also under investigation.

 We finish by noticing that, despite of our results are  indeed fully  compatible with the conclusions of \cite{skardal2014}, in particular with the fact that   anti-correlation between the frequencies of neighbor nodes 
 favors synchronization (larger values of ${\cal L}$), the role played by the interlayer connection symmetries, as discussed in Section III, allows us to interpret our results as
 a genuine manifestation of the dynamical phenomenon of  AISync
 in multilayer random networks with identical layers.

\section*{Acknowledgements}
The author  thanks FAPESP (grant 2013/09357-9) and CNPq for the financial support, and
R.S. Pinto, O. Saa, J. Climaco, A. Motter, and Y. Zhang for enlightening discussions. The core of  
our numerical computations was done by using the  NetworkX \cite{NetworkX} and SciPy \cite{SciPy}  
packages    for python.

\end{document}